\documentclass[vecphys,natbib]{svmult}

\usepackage{makeidx}         
\usepackage{graphicx}        
\usepackage{multicol}        
\usepackage[bottom]{footmisc}

\makeindex             

\usepackage{wjh-defs}
\usepackage{astrojournals}
\bibpunct{(}{)}{;}{a}{}{,}      

\setkeys{Gin}{width=\linewidth, height=0.8\textheight,
  keepaspectratio=true} 
\DeclareRobustCommand\HIInew{H\,\scalebox{0.9}[0.85]{II}}
\renewcommand\HII{\HIInew}
\def\Grad{\mathop{\vec{\nabla}}\nolimits}
\def\Div{\vec{\nabla} \!\cdot\!}

\DeclareRobustCommand\lesssim{\mathrel{\hbox{\footnotesize\rlap{\hbox{\lower3pt\hbox{$\sim$}}}\raise1.5pt\hbox{$<$}}}}

\newcommand\ThisVolume[1]{the chapter by #1 chapter in this volume}

\begin{document}

\title*{How to move ionized gas: an introduction to the dynamics of
  \HIInew{} regions}
\titlerunning{\HIInew{} region dynamics}
\author{W. J. Henney}
\institute{Centro de Radioastronom\'{\i}a y Astrof\'{\i}sica, UNAM Campus
  Morelia, Apartado~Postal 3-72, 58090~Morelia, Michoac\'an, M\'exico
  \texttt{w.henney@astrosmo.unam.mx}}
%
%
\maketitle

%

\HII{} regions are volumes of gas surrounding high-mass stars that are
ionized and heated by the stellar ultraviolet radiation. This
ionization and heating significantly raises the thermal pressure of
the gas, which is the principal driver of the dynamics of all but the
very smallest and very largest \HII{} regions. The simplest model of
\HII{} region evolution predicts a slow and decelerating expansion of
the ionized gas with very little internal motion, but this is rarely
observed. Instead, when examined in detail, such regions are found to
be highly structured with complex internal motions.

For reasons of space, this review is rather narrowly focused on the
traditional \HII{} regions that are found in our galaxy, ionized by
one or a handful of O~stars.  The dynamics of photoionized gas has a
much broader domain of application than this, covering such objects as
planetary nebulae, novae, Wolf Rayet nebulae, broad and narrow line
regions of active galaxies, and the reionization of the universe by
the first generation of stars. Many of these are covered in other
chapters in this volume. The physical priniciples presented here will
still apply in these broader contexts, but care must be taken, as some
of the shortcuts and approximations commonly used for \HII{} regions
may no longer be valid.

The chapter is divided into three sections. The first section
introduces the equations governing the dynamics and physical structure
of \HII{} regions and discusses the approximations that are commonly
employed. The second section presents the broad physical concepts that
provide the building blocks for constructing global models of \HII{}
regions. The third section provides an overview of such models, as
applied to the different evolutionary stages of \HII{} regions, with
particular emphasis on recent models of the evolution in a clumpy,
turbulent medium. 

\section{The equations}
\label{sec:equations}

The principal equations necessary for calculating \HII{} region
dynamics under the standard assumptions used in the field can be
divided into three groups. First, the Euler equations, which describe
the motion of the gas. Second, the radiative transfer equation, which
describes the emission and absorption of photons. Third, the rate
equations, which describe the transitions between different
atomic/ionic/molecular states. For each group, I first present the
general form of the equations, in which many difficult bits of physics
are hidden away in innocuous looking terms, before discussing the
sorts of approximations that might be made in different types of
models (all attempts to theoretically model \HII{} regions involve at
least \textit{some} level of approximation). 

\subsection{Euler equations}


These describe the motion of a non-viscous, non-relativistic gas of
density $\rho$, pressure $P$, and vector velocity $\vec{u}$, all of
which are functions of position, $\vec{r}$, and time, $t$. See, for
example, \citet{1992phas.book.....S} for full derivations. The ratio
of specific heats is assumed to be $\gamma = 5/3$, as is appropriate
for a monatomic gas. The equations are given here in there
\textit{conservation form}. Conservation of mass gives
\begin{equation}
  \label{eq:continuity}
  \frac{d\rho}{d t} + \Div \left(\rho \vec{u}\right) = 0  . 
\end{equation}
Conservation of momentum gives 
\begin{equation}
  \label{eq:momentum}
  \frac{d}{d t} (\rho \vec{u} ) + \Grad  \left(P + \rho u^2\right)
  = \vec{g} \rho ,  
\end{equation}
where $\vec{g}$ is the acceleration of the gas due to ``body forces''
caused by gravity, magnetic fields, or radiation
pressure. Conservation of energy gives 
\begin{equation}
  \label{eq:energy}
  \frac{d}{d t} \left( \frac32 P + \frac12 \rho u^2\right) 
  + \Div  \left[ \vec{u} \left(
  \frac52 P + \frac12 \rho u^2\right) \right]
  = \vec{u} \cdot \vec{g} \rho + H - C ,  
\end{equation}
where $H$ and $C$ are the volumetric rates of gas heating and cooling,
respectively, due to atomic processes. These equations are
supplemented by the ideal-gas equation of state, which defines the gas
temperature, $T$:
\begin{equation}
  \label{eq:eos}
  P = \frac{\rho k_\mathrm{B} T}{\mu m_\mathrm{H}}, 
\end{equation}
where $k_\mathrm{B}$ is Boltzmann's constant, $m_\mathrm{H}$ is the
mass of a hydrogen nucleus, and $\mu$ is the dimensionless
mean-mass-per-particle (for solar abundances, $\mu \simeq 1.3$ for
neutral gas and $\mu \simeq 0.6$--$0.7$ for ionized gas, depending on
the ionization state of helium). 

Although the above \textit{single-fluid} treatment is generally
adequate for the gas, it is not always such a good approximation when
dust grains are considered in detail. In such a case, one can employ a
multifluid treatment \citep{1979A&A....76..158G, 1979A&A....77..165G},
in which one uses a separate momentum equation for each fluid,
including explicit interaction terms for collisions between particles
of the different fluids.

\subsection{Radiative transfer}

The fundamental equation of radiative transfer describes the behavior
of the \textit{specific intensity}, $I_\nu(\vec{\hat{n}}, \vec{r})$
\citep{1978stat.book.....M}, which is a function of frequency, $\nu$,
position, $\vec{r}$, and direction, $\vec{\hat{n}}$:
\begin{equation}
  \label{eq:radtransfer}
  \frac{1}{c} \frac{d I_\nu}{d t} + \vec{\hat{n}} \cdot \Grad I_\nu  = 
  \eta_\nu - \chi_\nu I_\nu , 
\end{equation}
where $c$ is the speed of light, and $\eta_\nu$, $\chi_\nu$ are
respectively the emissivity and absorption coefficient, which may
contain an arbitrary amount of physics. For many purposes, it is
sufficient to work with angle-averaged moments of the specific
intensity, such as the \textit{mean intensity},
\begin{equation}
  \label{eq:mean-intensity}
  4 \pi J_\nu = \int_{4\pi} \!\!I_\nu \, d \Omega, 
\end{equation}
and the radiative flux, 
\begin{equation}
  \label{eq:flux}
  \vec{F}_\nu = \int_{4\pi} \!\!\vec{\hat{n}} I_\nu \, d \Omega. 
\end{equation}

\subsection{Rate equations} 

The general equation for the evolution of the partial density of a
particular state, $i$, can be written as
\begin{equation}
  \label{eq:rates}
  \frac{d n_i}{d t} + \Div (n_i \vec{u}) = 
  G_i + \sum_{j\ne i} R_{j\rightarrow i} n_j  - 
  n_i \left(S_i +  \sum_{j\ne i} R_{i\rightarrow j}\right) , 
\end{equation}
where $R_{i\rightarrow j}$, $R_{j\rightarrow i}$ represent the rates
of transitions between state $i$ and other states $j$, whereas $G_i$,
$S_i$ represent respectively sources and sinks of state $i$ due to
other processes. This form of the rate equation can apply to internal
states of an atom, an ion, a molecule, or a dust grain. Equally, it
may apply to the the total abundance of a given ion stage. In general,
the rates may be functions of the local radiation field, electron
density, temperature, or indeed any other physical variable. 

\subsection{How to avoid the dynamics}
\label{sec:avoiding-dynamics}

The most drastic simplification from the point of view of the dynamics
is to assume that there is no dynamics at all. This is the
\textit{static} approximation, which consists in setting $\vec{u} = 0$
and $dX/dt = 0$ for all quantities $X$. In this case,
equation~(\ref{eq:continuity}) is trivially satisfied, while
equation~(\ref{eq:momentum}) reduces to hydrostatic equilibrium, or
$P=\mathrm{constant}$ if there are no external forces. The energy
equation~(\ref{eq:energy}) reduces to $H = C$ and the rate equations
(\ref{eq:rates}) reduce to $\mathrm{sources} = \mathrm{sinks}$.  This
approximation is commonly employed in \textit{photoionization codes},
which treat the microphysics of the energy and rate equations in great
detail, and usually the radiative transfer also. As is shown below,
this approximation can be an acceptable one for the ionization and
thermal balance of the interior of an \HII{} region in the many
instances where the ionization/recombination and heating/cooling
timescales are much shorter than the dynamic timescale.

Slightly more realistic is the \textit{steady-state} or
\textit{stationary} approximation, in which one allows for non-zero
gas velocities, but still ignores all derivatives with respect to
time. This adds the \textit{advective} terms, of the form $\Div
(\vec{u} X)$, to the static form of the equations. These terms
represent the material flow of the quantity $X$ from one point to
another and are most important in regions of strong gradients in
$\vec{u}$ or $X$. In many instances, the flow timescale through the
part of the \HII{} region of interest is short compared with the
timescale for changes in the parameters of the flow, due, for
instance, to changes in the incident ionizing flux or the neutral
density outside the ionization front. In such cases, the stationary
approximation is often a very good one, provided that a suitable
reference frame is chosen.  In other cases, especially when the
long-term, global evolution of the \HII{} region is of interest, or if
dynamical instabilities are expected, then it is necessary to include
the non-steady $dX/dt$ terms and solve the \textit{fully
  time-dependent} equations. Even in this case, it is usually
sufficient to treat the radiative transfer equation in the stationary
limit, except when light-travel times are significant compared with other
timescales of interest \citep{2005astro.ph..7677S}. 

\subsection{How to avoid the atomic physics}
\label{sec:avoiding-physics}

In other contexts, one may wish to study the dynamics in detail
without worrying unduly about the finer points of the microphysics or
radiative transfer. To that end, various rules of thumb have evolved,
which give satisfactory approximations in many common
situations. However, the use of these approximations should be
justified on a case-by-case basis. The simplifications presented here
are at the extreme end of what one can get away with in a toy
model. For more serious applications, one should consult a text such
as \citet{2006agna.book.....O} for extra ingredients to add. 

\subsubsection{Thermal balance}
\label{sec:thermal-balance}

The temperature in galactic \HII{} regions is determined principally
by the balance between photoelectric heating (ejection of energetic
electrons by photoionization) and cooling due to forbidden lines of
the ions of heavy elements, which are excited by electron
collisions. The resultant equilibrium temperature is $\simeq 9000$~K
and rarely varies by more than 50\% throughout the region. It is
therefore natural to use an \textit{isothermal approximation} for the
ionized gas. 

Strong dynamic effects can cause this approximation to break down, but
this is rare in \HII{} regions. The dynamical terms in
equation~(\ref{eq:energy}) will not be greatly important unless the
dynamic timescale is shorter than the heating/cooling timescale, which
is typically 3--10 times shorter than the recombination
timescale. This requires very high gas velocities of $> 500\mathrm{\
  km\ s^{-1}}$ unless the ionization parameter (see below) is much
lower than is typical found.

\subsubsection{Transfer of ionizing photons}
\label{sec:transf-ioniz-phot}

Since hydrogen is (usually) the most abundant element, it is natural
to consider a hydrogen-only approximation, in which the only photons
of interest are those capable of ionizing hydrogen from its ground
state and the only opacity source considered for those photons is the
photoionization process itself. If the diffuse field is treated in the
on-the-spot approximation (see \citealp{2006agna.book.....O},
sec.~2.3), then only the direct stellar radiation need be explicitly
considered. Assuming the presence of only one ionizing star, then the
radiation field is monodirectional, so that all angular moments of the
radiation field are equal in magnitude, and in particular $4\pi J_\nu
= F_\nu = \left\vert\vec{F}_\nu\right\vert$. In this case, one can get
away with considering only the frequency-integrated flux of ionizing
photons: $f = \int_{\raise.2ex\hbox{$\scriptstyle\nu_0$}}^{\infty}
(F_\nu/h\nu) \, d\nu,$ where $\nu_0$ is the frequency corresponding to
the Lyman limit.
The radiative transfer equation then reduces to $\Div f = -\sigma
n_\mathrm{n} \vec{\hat{r}}$, where $ \vec{\hat{r}}$ is the radial
direction from the star, $n_\mathrm{n}$ is the number density of
neutral hydrogen atoms and $\sigma$ is the frequency-averaged
photoionization cross-section, weighted by the local ionizing
spectrum. In this approximation, $\sigma$ is a function only of $\tau
= \int \sigma n_\mathrm{n}\, dr$, and can be precomputed for a given
stellar spectrum.  In reality, dust opacity will often make a
significant contribution \citep[e.g.,][]{2004ApJ...608..282A,
  2005MNRAS.362.1038E} and should be included in any realistic
treatment.

\subsubsection{Ionization balance}
\label{sec:ionization-balance}

Collisional ionization of hydrogen is unimportant for $T < 20,000$~K
and three-body recombinations are negligible at typical \HII{} region
densities. The hydrogen ionization balance is then simply between
photoionization and radiative recombination, which in the
approximation discussed in the previous section is 
\begin{equation}
  \label{eq:ionization}
  \frac{d n_\mathrm{p}}{dt} + \Div (n_\mathrm{p} \vec{u}) =  
  n_\mathrm{n}  \sigma f
  - \alpha n_\mathrm{p} n_\mathrm{e}, 
\end{equation}
where $n_\mathrm{p}$, $n_\mathrm{e}$ ($\simeq n_\mathrm{p}$), and
$n_\mathrm{n}$ are the number densities of ionized hydrogen,
electrons, and neutral hydrogen, respectively, and $\alpha$ is the
appropriate recombination coefficient (Case B if the on-the-spot
approximation is used). If the radiative transfer is treated in more
detail, then $\sigma f$ should be replaced by $\int_0^\infty \!\!\!
\sigma_\nu (4 \pi J_\nu / h\nu) \,d\nu$.

 

\section{Physical concepts}

Before considering models for \HII{} regions as a whole, it is useful
to break down the problem into distinct physical ingredients, which
can be studied separately.

\subsection{Static photoionization equilibrium}
\label{sec:ioniz-stat-unif}

In order to compare and contrast with the results of dynamical models,
it is instructive to first consider the static case. The algebra is
simplest if one considers a uniform, plane-parallel slab, of density
$n$, one face of which illuminated by a perpendicular ionizing flux,
$f_0$, and in which all the extreme simplifying assumptions of the
previous sections are assumed to hold. The degree of ionization is
defined as $x = n_\mathrm{p}/n$ and, for reasonable values of $f_0$
and $n$, this fraction is very close to unity at the illuminated face:
$1-x_0 \simeq \alpha n/(\sigma f_0) \simeq 3 \times 10^{-6}
\,\varUpsilon^{-1}$, where $\varUpsilon = f_0 / (c n)$ is the
\textit{ionization parameter}, or ratio of photon density to particle
density, which typically takes values from $10^{-4}$ to $10^{-2}$. As
one moves to greater depths, $z$, in the slab, then integration of the
ionization balance and radiative transfer equations shows that the
ionizing flux decreases approximately linearly with depth, $f(z)
\simeq f_0 (1 - z/z_0)$, until one reaches a depth $z_0 \simeq f_0 /
(\alpha n^2)$, where the ionization fraction swiftly falls from
$x\simeq1$ to $x\simeq 0$ over a short distance, $\delta z \simeq 1/(n
\sigma )$. This depth to the ionization front, $z_0$, is referred to as the
\textit{Str\"omgren distance} and it is apparent that $\delta z / z_0
\simeq (1 - x_0)$, confirming that the front is indeed thin. The
ionization parameter, $\varUpsilon$, is the single most important
parameter describing the \HII{} region; as well as determining the
degree of ionization and the relative thickness of the  ionization
front, it is also proportional to the column density through the
ionized gas: $n z_0 = f_0 / (\alpha n) \simeq 10^{23} \,\varUpsilon
\mathrm{\ cm^{-2}}$. 

This last result allows one to see how the ionized region in
approximate photoionization equilibrium must respond to a change in
density. If the density in the slab increases, perhaps as the result
of compression by a shock, then the ionization parameter decreases and
so must the column of ionized gas, meaning that some of the gas must
recombine. Contrariwise, if the density decreases, perhaps due to
expansion,\footnote{The same holds for expansion in a spherical
  geometry since $n$ ($\sim r^{-3}$) falls more rapidly than $f$
  ($\sim r^{-2}$).} then the ionization parameter and the ionized
column will increase, leading to the advance of the ionization front
into previously neutral gas.

\subsection{Ionization front propagation}
\label{sec:dynam-ioniz-fronts}

The speed of propagation, $U$, of a plane, steady ionization front can
be determined by considering the jump conditions in the physical
variables across it \citep{1954BAN....12..187K}. Variables on the
neutral side of the front are given a subscript `1', and those on the
ionized side a subscript `2'. The velocity in the frame of reference
of the ionization front of the neutral gas, $v_1$, is equal in
magnitude to $U$, but opposite in direction. It is apparent that one
has $f_1 \simeq 0$ and $(1-x_2) \simeq 0$, so that by integrating
equation~(\ref{eq:rates}) across the front\footnote{In this
  approximation, recombinations in the front are ignored.} one has
$v_1 = f_2 / n_1$. When one then considers mass and momentum
conservation across the front, assuming isothermal sound speeds of
$a_1$ and $a_2$ on the two sides, one finds two classes of solution
(for example, \citealp{1992phas.book.....S}). Fronts with $v >
v_\mathrm{R} \simeq 2 a_2 \simeq 20 \mathrm{\ km\ s^{-1}}$ move
supersonically into the neutral gas and are called R-type (for
``rare''). Fronts with $v < v_\mathrm{D} \simeq a_1^2 / (2 a_2) \simeq
0.05 \mathrm{\ km\ s^{-1}}$ move subsonically into the neutral gas and
are called D-type (for ``dense''). If conditions are such that $f_2 /
n_1$ falls between $v_\mathrm{D}$ and $v_\mathrm{R}$, then the
ionization front will be preceded by a shock, which compresses the
neutral gas to a higher density, $n_1'$, so that $f_2 / n_1' <
v_\mathrm{D}$ and the ionization front becomes D-type. Fronts can be
further divided into those that contain an internal sonic transition,
which are termed strong, and those that do not, which are termed
weak. The only stable R-type fronts are weak, in which the front moves
supersonically with respect to both the neutral and the ionized gas
and such fronts show a low density contrast between the two sides: $1
\le n_1/n_2 \le 2$. The limit of extreme weak D-type fronts as $v_1
\rightarrow 0$ corresponds to the static case considered above, in
which the gas pressure is constant across the front so that $n_1/n_2 =
a_2^2 / a_1^2 \simeq 100$. When $v_1 = v_\mathrm{D}$, the front is
termed D-critical, with an ionized velocity that is exactly sonic with
respect to the front, $v_2 = a_2$, and an even higher density
contrast, $n_1/n_2 = 2 a_2^2 / a_1^2 \simeq 200$. Strong D-type fronts
can only occur if the sound speed has a maximum $a_\mathrm{m} > a_2$
inside the front and give $a_\mathrm{m} < v_2 < 2 a_\mathrm{m}$. They
are probably most relevant at very high metallicities, where the
equilibrium temperature of fully ionized gas can be much lower than
that of partially ionized gas.

When a volume of gas is initially exposed to ionizing radiation, the
flux is usually high enough that the ionization-front is R-type and
propagates supersonically through the gas. However, as the front
progresses to greater distances, an increasing proportion of the flux
is used up in balancing recombinations in the ionized gas. After a
time of order $1/(\alpha n)$ and a distance of roughly one Str\"omgren
distance, $f_2$ has become low enough that $U < 2 a_2$ and a preceding
shock detaches from the ionization front, which becomes D-type. Soon
after this point, the speed of the ionization/shock front falls below
the sound speed in the ionized gas, so that the subsequent evolution
of the front becomes sensitive to the internal conditions in the
\HII{} region, which determine the boundary conditions on the ionized
side of the front. In particular, if the ionized gas is free to flow
away from the front, then the front is liable to remain approximately
D-critical.  On the other hand, if the ionized gas is confined,
perhaps by a closed geometry (nowhere for the gas to go to) or a high
imposed pressure (e.g., from a stellar wind), then the front will be
weak~D. The latter case corresponds to the classical evolution of a
Str\"omgren sphere, which is described in many textbooks
\citep[e.g.,][]{1997pism.book.....D}.

The relative importance within the front itself of dynamic terms in
the ionization balance is large, of order $ u_2 \sigma / \alpha \simeq
10 M$, where $M = u_2/a_2$ is the Mach number of the ionized gas ($M =
1$ for a D-critical front). However, given that the front is generally
thin compared with the \HII{} region as a whole
(section~\ref{sec:ioniz-stat-unif}), the relative importance,
$\lambda_\mathrm{ad}$, of dynamics in the \textit{global} ionization
balance is much smaller, $\lambda_\mathrm{ad} = u_2/ (\varUpsilon c)
\simeq 0.003$--$0.3$. This is why photoionization models that ignore
dynamics are often a reasonable approximation for the interior of an
\HII{} region.

\subsection{Structure of a D-type front}
\label{sec:structure-d-type}

\begin{figure}
  \includegraphics{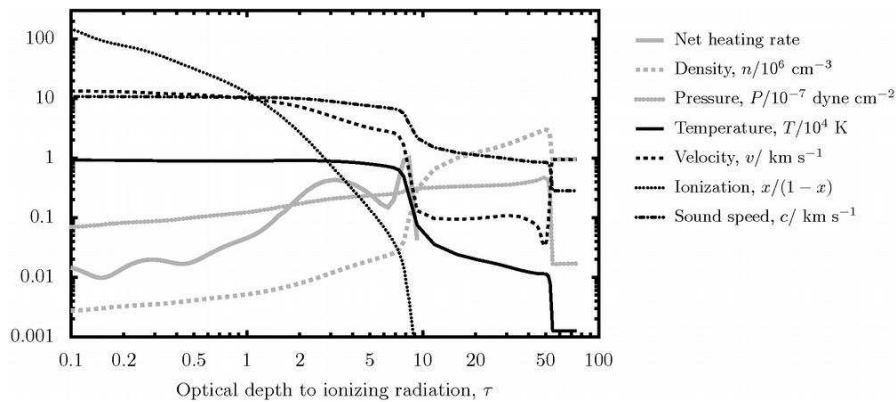}
  \caption{Physical variables in a typical D-critical ionization front
    taken from a cut along the axis of a two-dimensional
    hydrodynamical simulation of a champagne flow. Stellar photons are
    moving from left to right, but the gas is moving from right to
    left. Figure adapted from \citet{2005ApJ...627..813H}.}
  \label{fig:structure}
\end{figure}

When examined in detail, an ionization front is found to have
considerable internal structure. An example of an approximately steady
D-critical front is shown in Fig.~\ref{fig:structure}, which results
from a radiation-hydrodynamical simulation that uses rather simplified
atomic physics and an artificially lowered ionization parameter in
order to resolve the ionization front on a two-dimensional grid
\citep{2005ApJ...627..813H}. However, the principal results compare
well with much more detailed one-dimensional calculations of weak-D
fronts using a state-of-the-art plasma physics code
\citep{2005ApJ...621..328H}. In order to emphasize the regions of the
front where interesting changes occur, the physical variables are
plotted as a function of the mean optical depth for ionizing radiation
from the star. The velocity is in the direction of lower optical
depths and is given relative to the ionization front, which is moving
at $0.95 \mathrm{\ km\ s^{-1}}$ away from the star.

The net rate at which the gas gains energy from atomic processes
(photoelectric heating minus radiative cooling; solid gray line) has
two separate peaks in the front. The deeper peak occurs at $\tau
\simeq 8$ and results in the heating of the gas from $100$~K to
$9000$~K and its acceleration from $0.1\mathrm{\ km\ s^{-1}}$ to
$3\mathrm{\ km\ s^{-1}}$, but leaves the gas still largely neutral ($x
\simeq 0.01$). The shallower, broader peak occurs at $\tau \simeq 3$
and results in the complete ionization of the gas and its acceleration
to $\simeq 10\mathrm{\ km\ s^{-1}}$, with very little change in
temperature.\footnote{This can be compared with a simple analytic
  estimate for the optical depth to the ionization front (see
  Sec.~\ref{sec:dynam-ioniz-fronts}): $\tau_\mathrm{if} = -\ln
  \lambda_\mathrm{ad} \simeq 10.3 + 2.3 \log_{10} \varUpsilon$, giving
  $\tau_\mathrm{if} \simeq 1$ to $6$ for ionization parameters in the
  range $\varUpsilon = 10^{-4}$--$10^{-2}$.}
The gas then passes through a sonic point at $\tau \simeq 1$, at which
point $x \simeq 0.95$, and continues accelerating gradually up to
$\simeq 30\mathrm{\ km\ s^{-1}}$ as it flows back past the star (not
shown). Although the far downstream velocity of the ionized gas is
supersonic, the front is still D-critical since there is no internal
maximum in the sound speed. Instead, it is the divergence of the gas
streamlines, as the geometry deviates from plane-parallel, that allow
the gas to continue accelerating once it is fully ionized and
isothermal.

The shock that precedes the ionization front is also visible, at depth
of $\tau \simeq 50$. The shock is propagating into the neutral gas
slightly faster than the ionization front velocity, so this part of
the model is not exactly time-steady, but rather the column of shocked
neutral material grows with time. This part of the structure is the
least well-modeled in the simulation, since the physics of the
photodissociation region is not treated adequately, but see
\citet{2005ApJ...623..917H, 2005astro.ph.11165H} for dynamical models
that do treat this properly.

\subsection{Photoablation flows}
\label{sec:photoablation-flows}

If the neutral density is similar in all directions from the ionizing
star, then one would expect the radius to the ionization front to be
also similar, which would lead to a global geometry for the front that
is closed, or concave. Despite this, the opposite case is of great
relevance to \HII{} region dynamics, in which the ionization front is
convex; that is, it curves away from the ionizing star. This can
arise, for example, if dense condensations exist in the neutral gas,
which slow down the propagation of the ionization front, causing it to
``wrap around'' the obstacle. If this situation persists, then a
steady flow of ionized gas will establish itself that accelerates away
from the convex front. The density in this ionized
\textit{photoablation flow} increases sharply towards its base, where
it may greatly exceed the mean density in the \HII{} region. Models
for such flows were originally developed in the context of the
photoevaporation of small, self-gravitating globules
\citep{1968Ap&SS...1..388D, 1969Fisica..41.172K} but have a much wider
range of application. By adopting a cylindrical, rather than spherical
geometry, similar models can be applied to cases where the ionization
front wraps around a dense filament, as in the Bright Bar in the Orion
Nebula, and they can even be extended to cases where the ionization
front is flat, or slightly concave, as in the case of champagne flows
(see section~\ref{sec:later-phas-comp} below). In all cases, the
structure of the ionized flow is similar, provided only that the gas
streamlines diverge and that any pressure imposed at the downstream
boundary is low compared with the thermal pressure of the ionized gas
at the ionization front. In order for a steady-state approach to be
valid, it is also necessary that the timescale for significant changes
in the configuration of the neutral gas be long compared with the
dynamic timescale of the ionized flow. This condition is easily
satisfied in some cases, such as the Orion proplyds, but only
approximately satisfied in others, such as the radiation-driven
implosion of a neutral globule \citep{1989ApJ...346..735B}.

The simplest models proceed by patching together the solution for a
plane-parallel D-critical ionization front (see previous sections)
with an isothermal wind solution for the fully ionized flow. This is
valid so long as the width of the ionization front is very small
compared with its radius of curvature, $r_\mathrm{c}$, which is almost
invariably the case. The gas leaves the ionization front at the sound
speed but accelerates rapidly (see Fig.~6 of
\citealp{2003RMxAC..15..175H}), so that the ionized density falls off
more quickly than it would in a constant velocity wind, approximately
mimicking an exponential profile. As the Mach number of the flow
increases, the acceleration gradually lessens and a terminal velocity
is reached when the isothermal approximation breaks down, which occurs
when the expansion and heating timescales becoming comparable. For a
spherical flow with $\lambda_\mathrm{ad} = 0.01$, this occurs at
velocity of $\simeq 40\mathrm{\ km\ s^{-1}}$ and a radius of $ \simeq
10 r_\mathrm{c}$.

Unless the ionization parameter is rather low, $\lambda_\mathrm{ad}$
will be small and the flows are \textit{recombination dominated},
which means that the ionized flow has a high optical depth to ionizing
radiation and is in approximate static ionization
balance.\footnote{The opposite \textit{advection dominated} case is
  generally only seen in flows with low surface brightness
  \citep{2001RMxAC..10...57H}, most notably the knots in the Helix
  planetary nebula \citep{2001ApJ...548..288L, 2005AJ....130..172O},
  in which a high fraction of incident photons reach the ionization
  front to ionize new gas.} The effective recombination thickness of
the flow\footnote{Defined as $n_0^2 h_\mathrm{eff} = \int n^2\, dr$,
  where $n$ is the ionized density with value $n_0$ at the ionization
  front.} is $h_\mathrm{eff} = 0.12 r_\mathrm{c}$ for a spherical
geometry or $h_\mathrm{eff} = 0.3 r_\mathrm{c}$ for a cylindrical
geometry, so long as the curvature is positive (convex) and
$r_\mathrm{c} \ll r_\star$, where $r_\star$ is the distance of the
front from the ionizing star. For the case where $r_\mathrm{c} >
r_\star$, one instead finds $h_\mathrm{eff} \simeq 0.2$--$0.7 r_\star$
(see Fig.~7 of \citealp{2005ApJ...627..813H}). In either case, the
entire flow structure is uniquely determined by the incident ionizing
flux and the curvature of the front. If $r_\mathrm{c} \lesssim 0.2
r_\star $, then the photoevaporation flow is not capable of evacuating
the region all the way back to the ionizing star, so the incident flux
may be reduced by recombinations in the interior of the \HII{} region.


\subsection{Other ingredients}
\label{sec:secondary-effects}

\subsubsection{Stellar Winds}
\label{sec:stellar-winds}

High mass stars invariably drive fast ($\simeq 1000\mathrm{\ km\
  s^{-1}}$) stellar winds \citep{1999isw..book.....L}, Wind mass loss
rates are still very uncertain due to the poorly understood effects of
clumping on observational diagnostics \citep{2006ApJ...637.1025F}, but
are probably in the range $10^{-9}$--$10^{-7} M_\odot \mathrm{\
  yr^{-1}}$ for main-sequence O~stars. The effect of a stellar wind on
the dynamics of an \HII{} region depends crucially on whether or not a
hot bubble of shocked stellar wind can persist, which determines
whether the shell of shocked \HII{} region gas is
\textit{energy-driven} or \textit{momentum-driven} (see
\ThisVolume{Arthur}). In the energy-driven case,
\citet{2001PASP..113..677C} show that the shell swept up by the
stellar wind can dominate the expansion energy of the region at early
evolutionary times and when the ambient density is $> 10^4\mathrm{\
  cm^{-3}}$ (see also \citealp{1977A&A....59..161D,
  1996ApJ...469..171G}). The wind-blown shell is also capable of
trapping the ionization front in some instances. In the
momentum-driven case, the wind never dominates the global energetics
of the region but can nevertheless have an important local influence
on the internal dynamics. One example is the bowshocks found around
proplyds in the inner Orion Nebula
\citep{2001ApJ...561..830G}. Various mechanisms may prevent the
formation of the hot shocked bubble that is required for the
energy-driven case, including thermal conduction
\citep[e.g.,][]{1987A&A...177..243D} or dynamical instabilities
\citep{1988MNRAS.235.1011B} at the contact discontinuity, or
mass-loading of the wind due to embedded photoablating clumps or
proplyds \citep{1995MNRAS.277..700D, 2002RMxAA..38...51G}. All these
can cause enhanced cooling in the shocked wind, leading to the loss of
its thermal pressure. Specific instances of the importance of winds in
the evolution of \HII{} regions are described in
section~\ref{sec:glob-models-nebul} below.

\subsubsection{Radiation pressure}
\label{sec:radiation-pressure}

The transfer of momentum between the radiation field and the gas can
sometimes affect the dynamics of \HII{} regions. This arises in three
different ways: first, through trapped resonance line radiation
(mainly Lyman $\ualpha$); second, through the momentum of stellar
photons transferred during the photoionization process; and, third,
indirectly through collisional coupling with dust grains that are
accelerated by the absorption of stellar radiation. 

Although the diffusion of resonance line photons can in principle
introduce non-local couplings between disparate parts of an \HII{}
region, this is severely hampered by the presence of dust absorption
\citep{1980ApJ...236..609H}, which allows a purely local approach to
be used if the medium is homogeneous on scales $< 0.1 /
\kappa_\mathrm{d}$, where $\kappa_\mathrm{d}$ is the dust absorption
coefficient. By this means, \citet{1998AJ....116..322H} showed that the
resonance line radiation pressure is proportional to the gas pressure,
making a contribution of only $\simeq 5\%$ for a standard
dust-gas-ratio.

The radiative acceleration due to absorption of ionizing photons of
mean energy $\langle h \nu \rangle$ is $g_\mathrm{rad} \simeq (\langle
h \nu\rangle/c)( f / \rho h_\mathrm{abs})$, where $h_\mathrm{abs}$ is
the thickness over which the photons are absorbed.  The acceleration
due to pressure gradients in a photoablation flow
(section~\ref{sec:photoablation-flows}) is $g_\mathrm{flow} \simeq a^2
/ h_\mathrm{eff}$. Comparing the two, and putting $h_\mathrm{abs} =
h_\mathrm{eff}$, one finds $g_\mathrm{rad} / g_\mathrm{flow} \simeq
\varUpsilon \langle h \nu\rangle / (2 k T) \sim 10 \varUpsilon$, so
that the pressure of the ionizing radiation is of only secondary
importance unless the ionization parameter is higher than is typically
found in \HII{} regions. Assuming that the gas and grains are
effectively coupled, one finds a similar result for the absorption of
radiation by dust, although this can become more important for cooler
stars, where the relative luminosity at non-ionizing wavelengths is
higher. Radiation pressure on dust can also be important in the inner
parts of \HII{} regions due to the increased flux close to the
ionizing source,\footnote{This is not the case for radiation pressure
  on the gas because of its dependence on the product of the flux and
  \emph{neutral} density, which is roughly constant throughout the
  ionized region.} where it may be responsible for central cavities in
some cases \citep{1967ApJ...147..965M, 2002ApJ...570..688I}.

\subsubsection{Magnetic fields}
\label{sec:magnetic-fields}

The magnetic field makes a significant, perhaps dominant, contribution
to the pressure both in molecular clouds and in the diffuse
ISM\@. \HII{} regions are generally over-pressured with respect to
their undisturbed surroundings, but are within a factor of two of
pressure equilibrium with the shocked neutral gas that surrounds them.
The dynamic importance of magnetic fields inside the \HII{} region
depends sensitively on how the field strength, $B$, responds to
compression in the shock and rarefaction at the ionization front. This
can be approximately characterized by an effective adiabatic index,
$\gamma_\mathrm{m}$, such that $B^2 \propto
\rho^{\gamma_\mathrm{m}}$. Possible values of this index are bounded
by $\gamma_\mathrm{m} = 0$ for compression/rarefaction parallel to the
field lines of an ordered field and $\gamma_\mathrm{m} = 2$ for
compression/rarefaction perpendicular to the field, whereas
$\gamma_\mathrm{m} \lesssim 1$ seems to be indicated by
observations. The order-of-magnitude increase in sound speed between
the ionized and neutral gas means that the plasma $\beta$-parameter
(ratio of gas pressure to magnetic pressure) will be between 5 times
(if $\gamma_\mathrm{m} = 0$) and 2000 times (if $\gamma_\mathrm{m} =
2$) higher in the \HII{} region than in the undisturbed neutral gas,
assuming the \HII{} region to be five times overpressured. Thus, it is
plausible that the magnetic field should play a much lesser role in
the ionized gas than in the neutral gas. Nevertheless, the magnetic
fields can still have dramatic effects on \HII{} region dynamics,
particularly for the ionization front, and magnetohydrodynamic models
of these are presented in detail in \ThisVolume{Williams}.

\subsubsection{Instabilities}
\label{sec:instabilities}

The dynamics of the photoionized gas will also be influenced by many
different kinds of instabilities (see \citealp{2003RMxAC..15..184W} for
an overview), which may contribute to the detailed structure observed
in \HII{} regions. Different modes of instability at the ionization
front are found, depending on whether the shell of shocked neutral gas
outside it is thick \citep{2002MNRAS.331..693W} or thin
\citep{1979ApJ...233..280G, 1996ApJ...469..171G}. Recombinations in
the fully ionized gas are found to damp the instabilities in some
\citep{1964ApJ...140..112A, 2005ApJ...621..803M} but not all
\citep{2002MNRAS.331..693W} cases. The interaction between streams of
gas inside the \HII{} region, such as stellar winds or photoablation
flows, provides opportunities for further instabilities (e.g.,
Rayleigh-Taylor, Kelvin-Helmholtz) and these may interact with
ionization front instabilities in complicated ways. An example is
shown in the simulations of \citet{2005astro.ph.11035A}, where the
fragmention of a wind-driven shell triggers shadowing instabilities of
the type discussed by \citet{1999MNRAS.310..789W}.





\section{\HII{} region evolution}
\label{sec:glob-models-nebul}

While the classical scenario for the expansion of an ionized
Str\"omgren sphere in a constant density medium has clear didactic
value, it not necessarily relevant to the evolution of real \HII{}
regions. Massive stars form in dense and highly inhomogeneous
molecular clouds and begin to emit ionizing photons while they are
still accreting mass, possibly via a disk (see \ThisVolume{Franco \&
  Hoare} for an overview of high-mass star formation). Various
mechanisms may act to prolong the duration of the early phases of
evolution, either by physical confinement or by providing a reservoir
of neutral gas. Eventually the \HII{} region may break out from the
molecular cloud and become optically visible, but its evolution will
still be strongly affected by the strong density gradients in its
environment. Since high-mass star formation is strongly clustered,
extended \HII{} regions will tend to be excited by multiple OB
stars. As the region's age begins to exceed the main-sequence lifetime
of the highest mass stars, then the powerful winds from evolved stars
and subsequent supernova explosions will have a dramatic effect on the
dynamics.

\subsection{Early phases: hypercompact and ultracompact regions}
\label{sec:early-phas-hyperc}

The smallest observed \HII{} regions are of size $\sim
10^{-3}$~parsec, and are classified as hypercompact regions
\citep{1997ApJ...486L.103C, 2004ApJ...605..285S}. At these small
sizes, the escape velocity at the ionization front is larger than the
ionized sound speed, and so gravitational effects are important. If
the central star is still accreting via a spherical Bondi-Hoyle flow,
then the accretion velocity just outside the ionization front exceeds
the R-critical velocity of $2c_\mathrm{i}$, so the gas suffers only a
mild deceleration in the front and continues its accretion onto the
star \citep{2002ApJ...580..980K}. During this phase, the ionization
front does not expand dynamically, but can only increase its radius if
the ionizing luminosity of the central star increases
\citep{2003ApJ...599.1196K}. This phase is therefore rather
long-lived, with a duration given by the timescale for the nascent
star to grow by accretion, of order $10^5$~years.

\HII{} regions with sizes $\sim 0.01$--$0.1$~parsec and densities $>
10^5\mathrm{\ cm^{-3}}$ are classified as ultracompact and these are
much more numerous and well-studied than the hypercompact regions
\citep[e.g.,][]{1989ApJS...69..831W, 1994ApJS...91..659K,
  2005ApJ...624L.101D}. They show a diverse range of morphologies,
such as cometary, shell-like, bipolar, or irregular, although roughly
half are spherical or unresolved. The high number of observed
ultracompact \HII{} regions in the galaxy has been taken to imply a
lifetime for this phase of $\simeq 10^6$~years
\citep{1989ApJS...69..831W}, although it should be noted that other
estimates are as short as $5\times 10^4$~years
\citep{1996A&A...314..776C}. A remarkable diversity of dynamical
models have been proposed in order to explain this lifetime, which is
longer than the time for a classical Str\"omgren sphere to expand past
$0.1$~parsec. In some models, the ionized gas is physically confined
to ultracompact sizes, either by the ram pressure of the ambient
medium as the star moves through the molecular gas
\citep{1990ApJ...353..570V}, or by an accretion flow onto the star
\citep{2002ApJ...580..980K, 2005ApJ...621..359G}, or merely by the
high static pressure (thermal plus magnetic plus turbulent) of the
molecular cloud core \citep{1995RMxAA..31...39D,
  1996ApJ...469..171G}. Other models allow the ionized gas to expand
freely but provide a reservoir of neutral gas, the photoablation of
which sustains the brightness of the ultracompact core: either an
accretion disk around the high-mass star \citep{1994ApJ...428..654H,
  1996A&A...315..555Y, 1997A&A...327..317R, 2004ApJ...614..807L}, or
dense neutral clumps \citep{1995MNRAS.277..700D, 1996ApJ...468..739L,
  1996MNRAS.280..661R}. In many instances, ultracompact \HII{} regions
are seen to be embedded in larger-scale, more diffuse emission
\citep{1999ApJ...514..232K}, which would tend to favor the second
class of models if the extended emission were physically connected
with the ultracompact region. However, it is entirely possible that
each of these models is applicable to some subclass of ultracompact
regions.

\subsection{Later phases: compact and extended regions}
\label{sec:later-phas-comp}

At least some ultracompact \HII{} regions eventually expand to larger
sizes, forming the class of compact \HII{} regions
\citep{1967ApJ...150L.157M}, with sizes $0.1$--$0.5$~parsec and
densities $\simeq 10^4\mathrm{\ cm^{-3}}$. Most compact regions are
embedded inside extended \HII{} regions, which are typically several
parsecs in size, with densities $\simeq 100\mathrm{\ cm^{-3}}$. The
distribution of morphological types is similar to that for
ultracompact regions \citep{1992ApJS...79..469H, 1993ApJS...86..475F},
although most modeling effort has gone into explaining the cometary
shaped regions. These are similar in appearance to many optically
visible \HII{} regions, which are classified as blister-type
\citep{1978A&A....70..769I}, and of which the best-known example is
the Orion Nebula \citep{2001ARA&A..39...99O}. The low extinction to
these optical regions proves that they must be on the near side of any
accompanying molecular gas, and they generally show blueshifted
velocities of order the ionized sound speed in optical emission lines,
indicative of flow away from the molecular cloud
\citep{1973ApJ...183..863Z}. A broad class of models, commonly
referred to as champagne flows, has been proposed to account for these
objects. These models share the property that strong density gradients
in the neutral/molecular gas allow the ionization front to break out
in some directions, leading to a flow of high-pressure ionized gas in
the same direction. The original model of \citet{1979A&A....71...59T}
considered the one-dimensional propagation of an ionization front
inside a dense cloud with a sharp edge. Once the ionization front
reaches the edge of the cloud, it rapidly propagates through the much
rarer intercloud medium and is followed by a strong shock that is
driven by the higher pressure ionized cloud material. A rarefaction
wave simultaneously travels back into the ionized cloud, initiating
the champagne flow that accelerates the ionized gas up to several
times the ionized sound speed as it flows away from the
cloud. Following studies extended this work to two dimensions (see
references in \citealp{1986ARA&A..24...49Y}, section 3.3) and
considered the effects of disk-shaped clouds and more gradual cloud
boundaries. 

\begin{figure}
  \includegraphics{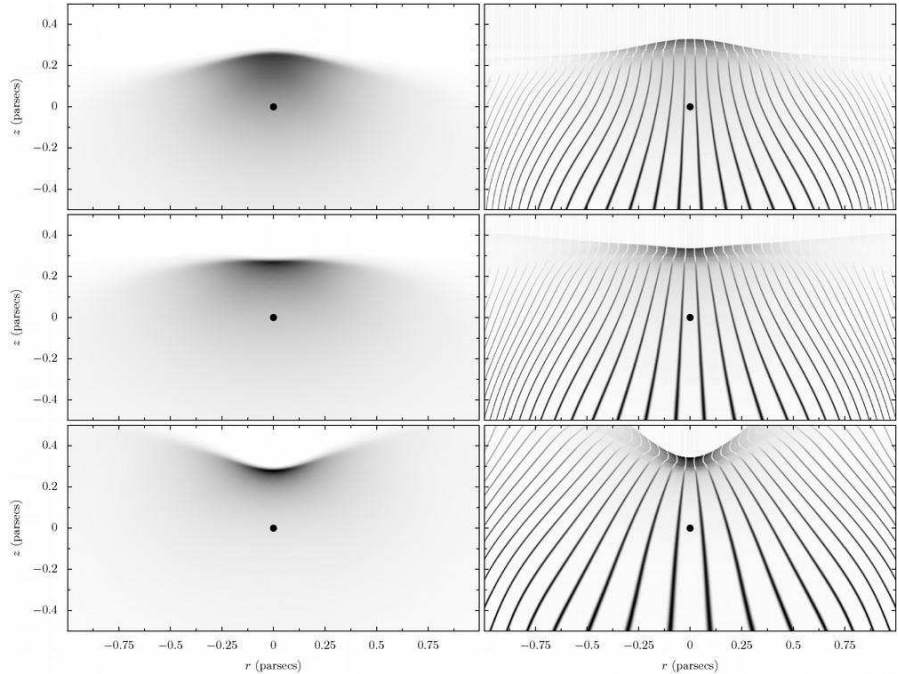}
  \caption{Simulations of quasi-steady champagne flows, produced by
    the ionization of the interface of a dense molecular cloud
    (adapted from \citealp{2005ApJ...627..813H}). Three different
    models (top to bottom) are shown with different lateral density
    profiles in the neutral gas. The left panels shows the ionized
    density, while the right panels show the total pressure with
    superimposed streamlines, the darkness of which indicates the gas
    velocity. The position of the ionizing star is shown by a black
    circle in each case. }
  \label{fig:champ}
\end{figure}

Early work on the dynamics of champagne flows considered scenarios
that were intrinsically non-steady. However, it is also possible to
construct quasi-stationary champagne flow models
\citep{2005ApJ...627..813H} in which the structure of the ionized flow
remains constant over several times its dynamical timescale. These
models are valid so long as the density of the neutral gas that
confines the ionization front on one side is high enough that the
ionization front moves slowly and encounters constant upstream
conditions during the evolution. It can be shown that in the
steady-state limit a divergence of the ionized flow is necessary in
order to produce significant acceleration. This is beacuse in a
strictly plane-parallel geometry, and in the absence of body forces,
the acceleration is proportional to the gradient of the sound speed,
which is almost zero in the isothermal ionized gas. Examples of such
flows with different degrees of divergence are shown in
Fig.~\ref{fig:champ}. The degree of divergence is controlled by the
neutral density profile in the lateral direction (perpendicular to the
sharp density gradient at the surface of the cloud). When this density
profile is constant, as in the upper panels of Fig.~\ref{fig:champ},
the ionization front is concave (negative curvature), leading to weak
divergence of the flow and slow acceleration. For a steep lateral
density profile, as in the lower panels of Fig.~\ref{fig:champ}, the
ionization front becomes convex (positive curvature), giving a strong,
almost spherical, divergence to the flow, which now accelerates much
more strongly. This configuration is similar to that seen in globule
flows. Approximate analytic calculations suggest that the ionization
front should be flat when the lateral density profile is proportional
to $1/(1+r^2)$, in which the distance, $r$, from the symmetry axis has
been scaled by the axial offset between the star and the ionization
front. Numerical simulations confirm that this is indeed the case, as
shown in the central panels of Fig.~\ref{fig:champ}.

\begin{figure}
  \setkeys{Gin}{width=0.47\linewidth}
  \includegraphics{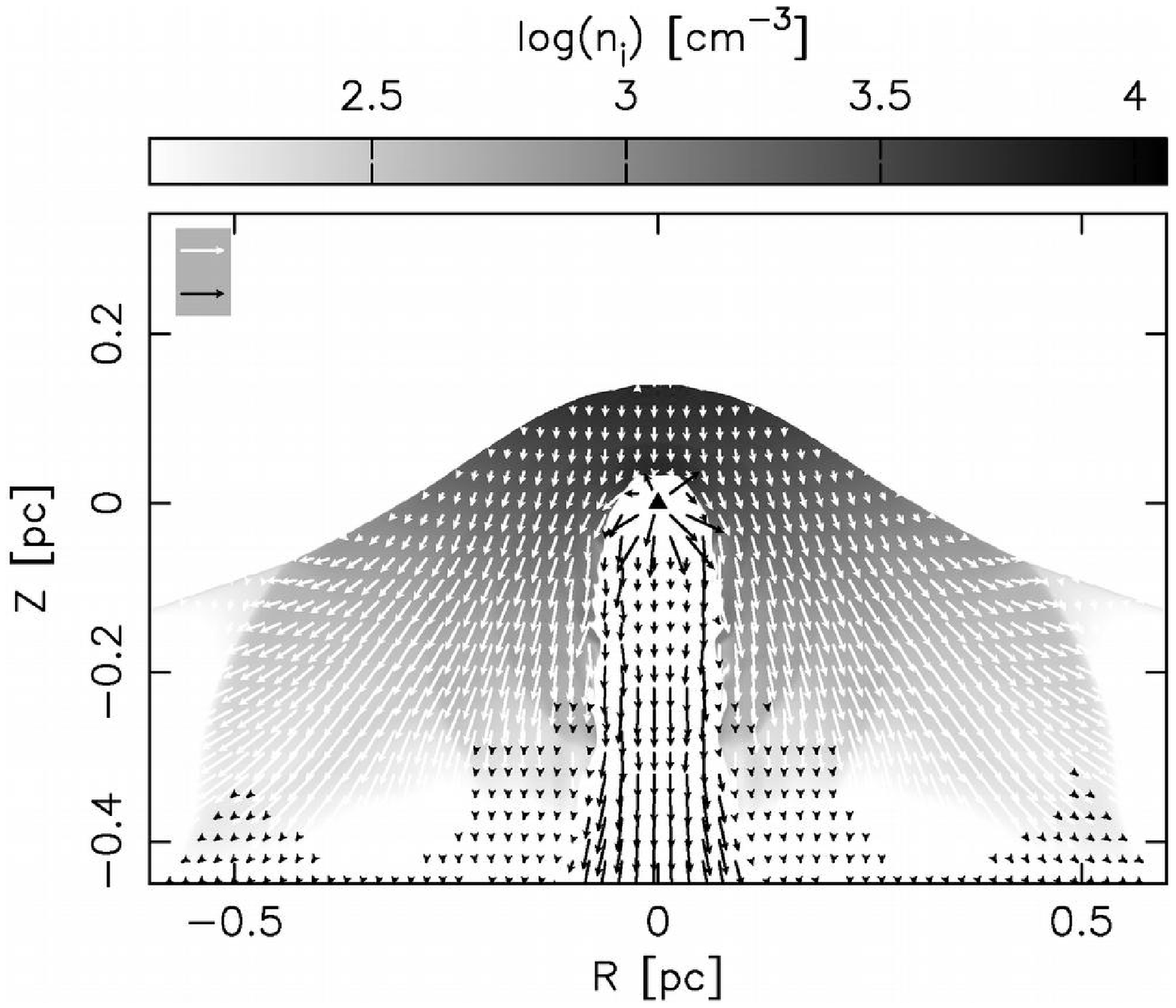}\hfill 
  \includegraphics{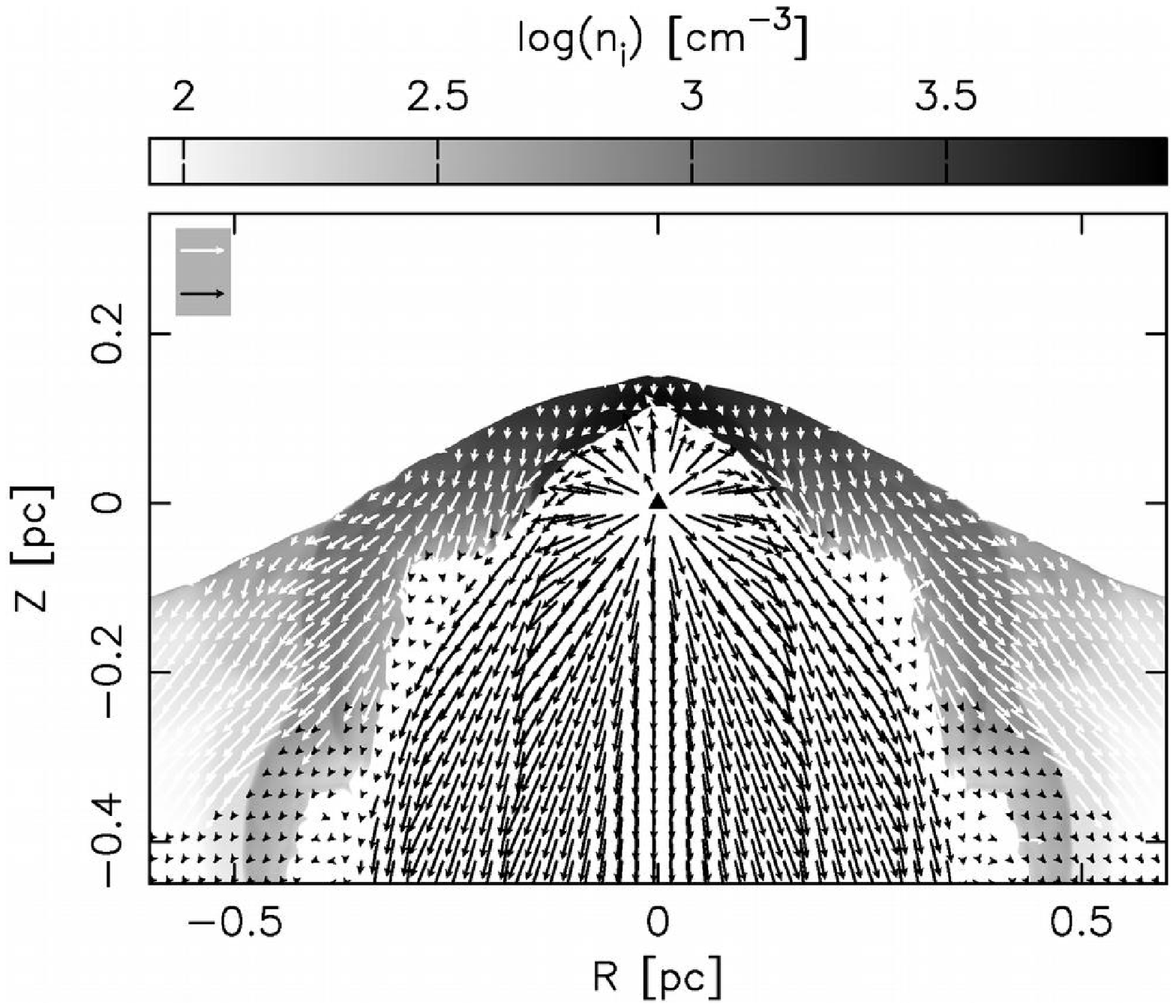}
  \caption{Simulations of the interaction of a champagne flow with a
    stellar wind (from \citealp{2005astro.ph.11035A}). Grayscale shows
    log of ionized density and arrows show gas velocity on two
    different scales: the longest white arrow represents $30\mathrm{\
      km\ s^{-1}}$ while the longest black arrow represents
    $2000\mathrm{\ km\ s^{-1}}$. The left panel shows a model with a
    weak wind (mass loss rate of $10^{-7}M_\odot \mathrm{\ yr^{-1}}$)
    and the right panel shows a model with a strong wind
    ($10^{-6} M_\odot \mathrm{\ yr^{-1}}$). Both models have a wind
    velocity of $2000\mathrm{\ km\ s^{-1}}$ and an
    ionizing photon luminosity of $2.2\times 10^{48} \mathrm{\
      s}^{-1}$. }
  \label{fig:jane}
\end{figure}

The stellar wind from the ionizing star may also be expected to have
an effect on the dynamics of the champagne flow and this was first
modelled by \citet{1997A&A...326.1195C}. An alternative explanation
for the cometary shapes shown by many regions is the bowshock model
\citet{1990ApJ...353..570V} previously discussed in the context of
ultracompact regions. In this model, the ionizing star moves
supersonically through the molecular gas and the ionization front is
trapped in the dense shocked shell formed by the interaction between
the stellar wind and the ambient gas. An extension of this scenario
was studied by \citet{2005astro.ph..8467F}, in which the ionizing star
moves in and out of a high density molecular cloud core. An exhaustive
study of the interplay between champagne flows, stellar winds, and
stellar motion was carried out by \citet{2005astro.ph.11035A}, who
found a great variation in the resultant morphology, depending on the
strength of the stellar wind. All the \citeauthor{2005astro.ph.11035A}
models have concave ionization fronts, as in the top row of
Fig.~\ref{fig:champ}. If the stellar wind is weak, as in the first
panel of Fig.~\ref{fig:jane}, then the champagne flow is hardly
affected, except in a narrow region around the axis, where it forms a
weak bowshock around the stellar wind. A stronger stellar wind, as in
the second panel of Fig.~\ref{fig:jane} has a much more dramatic
effect. The point of pressure balance along the axis between the
stellar wind and the \HII{} region gas is pushed closer to the
ionization front than where the sonic point would have been in the
champagne flow. As a result, there is no champagne flow on the axis,
although away from the axis a transonic champagne flow does develop,
which then shocks against the stellar wind. The large obstacle
provided by the wind means that the champagne flow acceleration is
predominantly parallel to the ionization front in this case. The
inclusion of a stellar velocity in the direction of increasing
gradient produces only small changes in the results unless the stellar
velocity is higher than the ionized sound speed. Such models always
show a greater resemblance to champagne flow models than to classical
bowshock models unless the density gradient in the neutral cloud is
very shallow. Either cometary or shell-like morphologies can be
produced, depending on the steepness of the neutral density
distribution.

Another class of models considers the radial expansion of an \HII{}
region in a spherical cloud with a density distribution steeper than
$r^{-3/2}$ \citep{1990ApJ...349..126F, 2002ApJ...580..969S,
  2003RMxAC..15..166L}, in which no static equilibrium solution exists
for the position of the ionization front.  When the star turns on, an
R-type ionization front quickly propagates to large radii, and the
ionized gas adopts a configuration in which a roughly uniform density
core, bounded by a shock, expands slightly faster than the ionized
sound speed. Such a region is doomed to rapidly decline in brightness
after a few sound crossing times. Confusingly, these entirely
density-bounded models are also referred to as champagne flows,
although they are significantly different from the models considered
in the previous paragraph. A classical champagne flow is
ionization-bounded on at least one side and will maintain a high
brightness indefinitely, as long as sufficient neutral gas exists to
confine the ionization front close to the ionizing star on that side.

At even larger scales, any expansion of the \HII{} region will stop
when it comes into pressure balance with the local interstellar
medium. The mean pressure in spiral arms at radius of solar circle is
$\simeq 5 \times 10^4~\mathrm{K\ cm^{-3}}$, which is 90\% non-thermal
\citep{2005ARA&A..43..337C}. This would balance that of the ionized
gas when its density has fallen to a few $\mathrm{cm^{-3}}$. The
equivalent Str\"omgren radius is $\simeq 30$~parsec for a typical
O~star, but would be would be larger for regions ionized by a cluster
of stars, in which case vertical density gradients in the disk may
become important. However, the time required for an \HII{} region to
expand to such sizes is comparable with the evolutionary timescales
for high-mass stars, so that the powerful winds from the later stages
of stellar evolution and the expansion of supernova remnants will have
a profound effect on the dynamics of the region. Similar
considerations apply to \HII{} regions seen in external galaxies
\citep{1990ARA&A..28..525S}, although the largest \HII{} regions ($>
500$~parsec) show highly supersonic velocities
\citep[e.g.,][]{1981MNRAS.195..839T, 1983ApJ...274..141G,
  2005A&A...431..235R}, which correlate with the luminosity of the
region and may be indicative of virial equilibrium of the ionized gas
in the gravitational potential of the star cluster.

\subsection{Clumps and turbulence}
\label{sec:clumps-inst-turb}

One shortcoming of the models mentioned in the previous section is
that they consider only smooth density distributions for the neutral
gas into which the \HII{} region propagates. In reality, this is
unlikely to be the case, since the molecular clouds in which high-mass
stars form are known to be highly structured, which is probably the
result of supersonic turbulence \citep{2004ARA&A..42..211E}. Dynamic
models that attempt to take into account the clumpy nature of the
neutral gas were first proposed by John Dyson and collaborators
\citep{1995MNRAS.277..700D, 1996MNRAS.280..667W, 1996MNRAS.280..661R,
  1998MNRAS.298...33R} in order to explain the morphology and
lifetimes of ultracompact \HII{} regions. In these models dense
neutral clumps are assumed to be distributed throughout the ionized
volume and to act as mass-loading sites for a stellar wind from the
ionizing star. The models are agnostic with respect to the physics of
the mass injection process, which is not modeled directly, but which
may be ablation by photoionization or hydrodynamic processes. Instead,
the flow is treated statistically under the assumption of prompt and
complete mixing of the ablated gas with the stellar wind.  Similar
models have also been presented that treat the ablation process in
more detail and explicitly allow for the finite lifetime of the clumps
\citep{1996ApJ...468..739L, 1997ApJ...484..810A}. Both sets of models
result in a shell-like morphology for the \HII{} region (if the clumps
are distributed uniformly), bounded by a recombination front through
which the ionized wind flows.

\begin{figure}
  \includegraphics{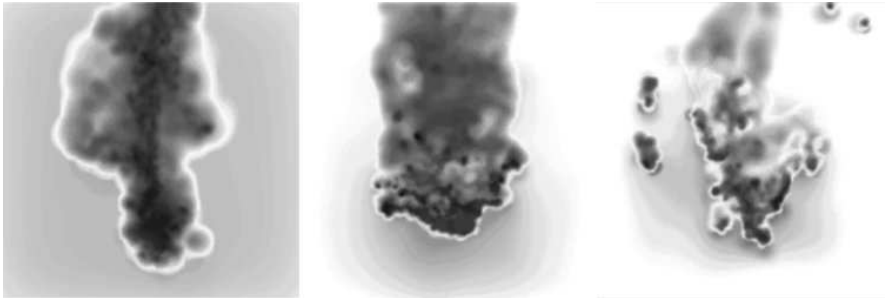}
  \caption{Three-dimensional simulation of the photoevaporation of a
    clumpy self-gravitating globule (adapted from
    \citealp{2003MNRAS.338..545K}). \textit{Left:} Phase of maximum
    compression, $1.6\times 10^5$~years. \textit{Center:} Equilibrium
    cometary phase, $2.8\times 10^5$~years. \textit{Right:}
    Fragmentaion phase, $4.2\times 10^5$~years. The grayscale measures
    the absolute value of the difference between the column density of
    the ionized and neutral gas (white corresponds to the ionization
    front). Box size is 0.85~parsec in the left panel and 1.8~parsec
    in the other panels.}
  \label{fig:kessel}
\end{figure}

A more detailed treatment of the interaction of ionizing photons with
a clumpy medium is possible only with three-dimensional numerical
simulations. The first such simulation to be carried out studied the
response of a static equilibrium self-gravitating spherical globule to
an external ionizing flux \citep{2003MNRAS.338..545K}. The density
distribution of the initial globule (of mass 40~$M_\odot$ and radius
$1$~parsec, on the margin of Jeans instability) is perturbed by
Gaussian fluctuations of amplitude $\simeq 50\%$. It is found that
the turbulence generated behind the shock driven into the globule by
the ionization front is sufficient to prevent gravitational collapse
of the globule at its point of maximum compression, despite collapse
having occurred at this point in a companion simulation of a globule
without density perturbations. After $\simeq 5\times 10^5$~years, the
neutral gas that has survived photoablation has broken up into smaller
sub-globules (Fig.~\ref{fig:kessel}). Similiar simulations have also
been carried out of the photoevaporation of much smaller,
non-self-gravitating, clumpy globules \citep{2005RMxAA..41..443G}. In
this case, although some fragmentation of the globule occurs, the
fragments remain physically attached to one another until the globule
is completely evaporated.

\begin{figure}
  \includegraphics{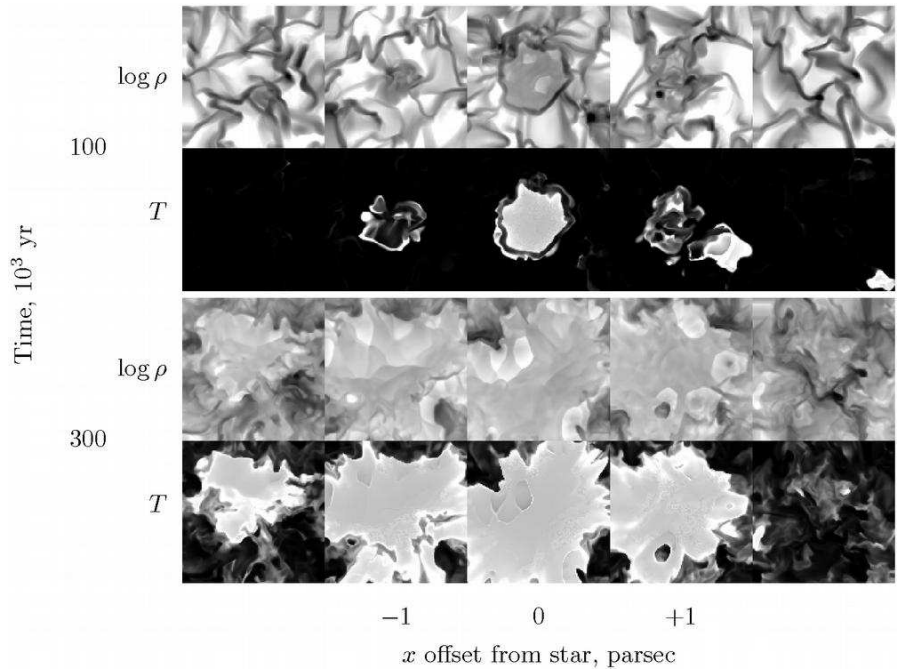}
  \caption{Dynamical evolution of an \HII{} region in a turbulent
    molecular cloud, adapted from \cite{2005astro.ph.12554M}. For each
    of two evolutionary times, 100,000 and 300,000~years, five $yz$
    slices through the $(4~\mathrm{parsec})^3$ computational box are
    shown, with offsets from the ionizing star as shown on the bottom
    axis. Density is shown on a negative logarithmic scale between 10
    (white) and $10^5~\mathrm{cm}^{−3}$ (black). Temperature is shown
    on a positive linear scale between 0 (black) and $10^4$~K (white). }
  \label{fig:turb}
\end{figure}

Simulations of hydrodynamical and magnetohydrodynamical turbulence in
molecular clouds \citep[e.g.,][]{1999ApJ...524..169M,
  2005ApJ...618..344V} have now reached a level of refinement that
makes them suitable to use as initial conditions for the evolution of
\HII{} regions. \citet{2004ApJ...610..339L} carried out a preliminary
investigation of this problem by studying only the initial R-type
propagation of the ionization front, before the dynamics of the
ionized gas becomes important. This approach was extended by
\citet{2005astro.ph.12554M}, who carried out a full
radiation-hydrodynamic simulation of the evolution of an \HII{} region
in a turbulent medium, following the birth of an $\simeq 25~M_\odot$
star inside the densest molecular clump ($\simeq
10^6~\mathrm{cm}^{-3}$) that formed in the \citet{2005ApJ...618..344V}
turbulence simulation. 

The results of this simulation are shown in Fig.~\ref{fig:turb} for
two evolutionary times. At the earlier time, the \HII{} region is
mainly confined to a compact core of radius $\simeq 1$~parsec,
although the ionization front has already broken out to the boundary
of the grid through a corridor of low-density neutral gas in one
direction, as can be seen in the lower-right corner of the rightmost
temperature image. The density variations in the ionized gas are
rather mild compared with those in the neutral gas and are of a
different nature. Instead of dense sheets and filaments, one finds an
almost constant density gas filling roughly half of the ionized
region, with the remainder occupied by low-density cavities carved out
by transonic photoablation flows, as described in
section~\ref{sec:photoablation-flows} (see also
\citealp{2001MNRAS.327..788W, 2003RMxAC..15..175H}). This structure of
the ionized gas can be better appreciated at the later evolutionary
time, where multiple photoablation flows are seen, streaming off dense
neutral condensations. The highest ionized densities are found at the
ionization fronts of these flows and their mutual collisions produce
many weak shocks, which generate density structure in the interior of
the \HII{} region. Some of these structures become compressed
sufficiently to significantly absorb the ionizing photons, causing the
ionized gas beyond them to temporarily recombine. Regions where this
has occured can be seen as intermediate-temperature gas in the bottom
row of Fig.~\ref{fig:turb}.  The photoablation flows help maintain a
high velocity dispersion of the ionized gas, which remains roughly
equal to the ionized sound speed during the entire evolution, even
though the net radial expansion velocity falls to much lower values
\citep[Fig.~4]{2005astro.ph.12554M}. This mechanism is similar to that
proposed by \citet{1968Ap&SS...1..388D} to explain the velocity
dispersion observed in the Orion Nebula.

\begin{figure}
  \includegraphics{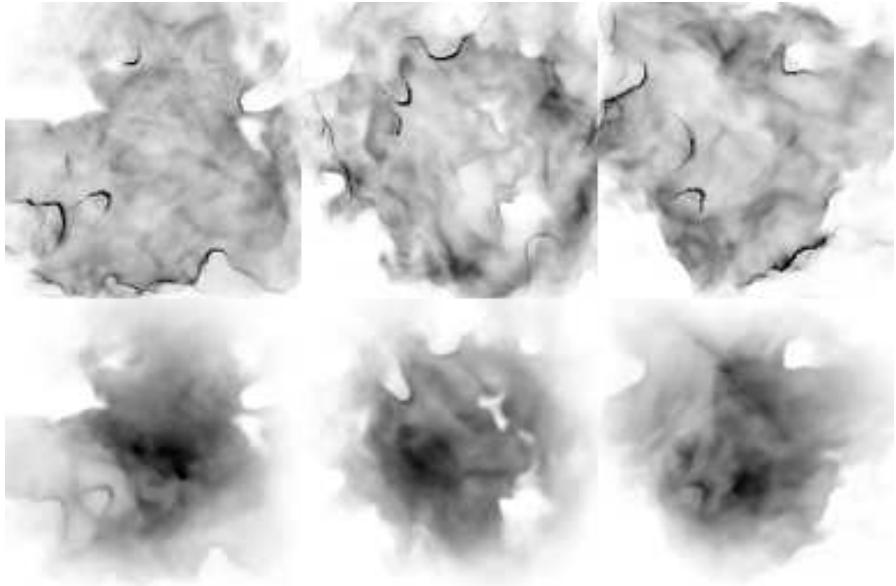}
  \caption[]{Fake emission line images from the simulation shown in
    Fig.~\ref{fig:turb} at an evolutionary time of
    400,000~years. \textit{Top row:} [\ion{N}{2}]
    6584~\AA{}. \textit{Bottom row:} [\ion{O}{3}] 5007~\AA{}. Views
    along the $x$, $y$, and $z$ axes are shown from left to right,
    respectively.}
  \label{fig:turb-emiss}
\end{figure}

\begin{figure}
  \includegraphics{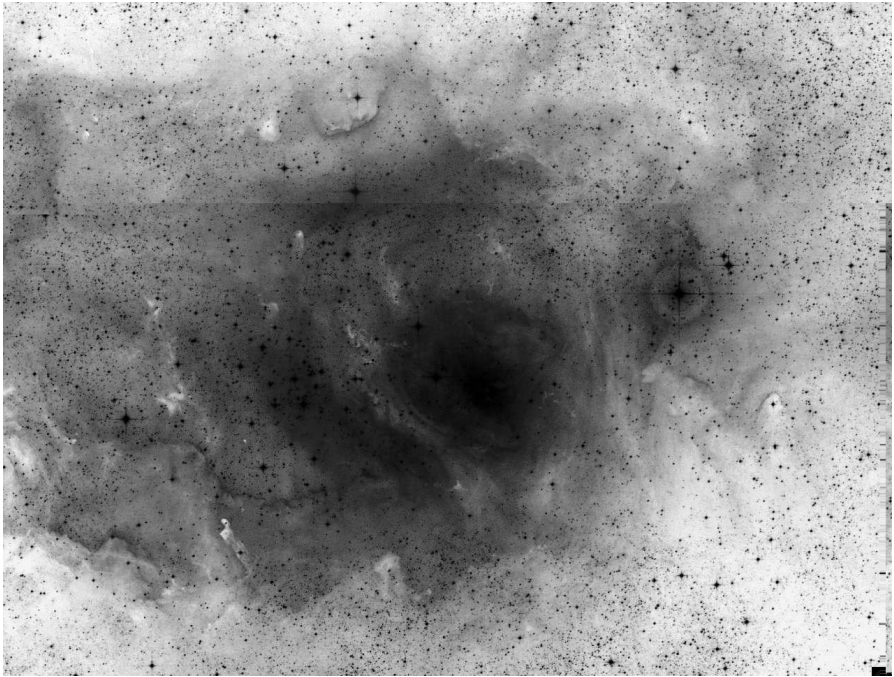}
  \caption{The Lagoon Nebula (M8). Combination of red and infrared
    continuum images from the Second Epoch Sky Survey of the UK Schmidt
    Telescope, operated by the Anglo-Australian Observatory, digitized
    by the Space Telescope Science Institute.  A highly non-linear
    grayscale mapping has been used to show both bright and faint
    structures at the same time. The field of view is
    $\simeq 27 \times 21$~parsec, assuming a distance of 1500~parsec.
  }
  \label{fig:lagoon}
\end{figure}

The predicted appearance of these simulations in optical emission
lines is shown in Fig.~\ref{fig:turb-emiss}. The [\ion{N}{2}] line
traces the low-ionization gas close to the ionization front, and these
images are dominated by the bright rims of photoablation flows, while
the [\ion{O}{3}] line traces the higher ionization gas in the interior
of the nebula, which shows a more diffuse aspect in the images. The
predicted emission line structure is very similar to that observed in
real nebulae, such as the Lagoon Nebula shown in
Fig.~\ref{fig:lagoon}. The broad-band filters used in that image
include optical emission lines of both low and high ionization, so it
should be compared with the superposition of the [\ion{N}{2}] and
[\ion{O}{3}] images (see Fig.~1 of \citealp{1994RMxAA..29...88D} for
pure emission line images of the inner regions of this nebula). As in
the simulations, one sees sharply bounded bright arcs and dark
extinction features in the low-ionization periphery of the nebula,
together with more diffuse emission in the higher ionization core. It
should be noted that the Lagoon Nebula is several times larger than
the simulation and is ionized by several massive stars, but a similar
morphology is seen in more compact nebulae, such as M16, M20, and M42.

Such sculpted structures at the boundary of the nebula can also be
produced via instabilities, even if the ambient neutral gas is
initially smoothly distributed \citep[e.g.,][]{1996ApJ...469..171G}. A
clear example of this is seen in Fig.~3 of
\citet{2003RMxAC..15..184W}, where large-scale density gradients lead
to a thin-shell instability and the formation of multiple
photoablation flows from the now clumpy shocked neutral layer. Whether
ionization front instabilities or pre-exisiting structures in the
neutral gas are more important in a given region is still an open
question. However, the density contrasts found in molecular clouds on
parsec scales are much more extreme than those seen in \HII{} regions,
due to the highly supersonic nature of the turbulence and the effects
of self-gravity. Thus, it seems likely that the structure of the
neutral gas is dominant in shaping \HII{} regions at these scales. At
smaller scales, the turbulent velocity dispersion, and hence the
density contrast, is much reduced, so that ionization front
instabilities may play a more important role, although these tend to
saturate at amplitudes of order the recombination length
\citep{2002MNRAS.331..693W}, which is approximately ten times the
ionization front thickness, or $\simeq n^{-1}$~parsec, where $n$ is
measured in $\mathrm{cm^{-3}}$.

\section*{Summary}
\label{sec:closing-comments}

\HII{} regions, particularly the extended, optically visible variety
have the reputation of being ``messy'' objects, ill-suited to the
austere theoretical prejudice of perfect symmetry. However, as I have
tried to illustrate in this review, even their messiest aspects are
beginning to fall under the purview of dynamic modelling.  We are
still some way from a totally satisfactory model of any particular
object, let alone \HII{} regions as a class, but a large amount of
progress has been made.

\bibliographystyle{astroads} 
\bibliography{All-Sorted,Extras}

\end{document}